\documentclass[12pt,preprint]{aastex}

\def\degree{\ifmmode {^\circ}\else {$^\circ$}\fi}
\def\rstar{\ifmmode {\, R_{\star}}\else $R_{\star}$\fi}
\def\msol{\ifmmode {\, M_{\odot}}\else $M_{\odot}$\fi}
\def\rsol{\ifmmode {\, R_{\odot}}\else $R_{\odot}$\fi}
\def\lsol{\ifmmode {\, L_{\odot}}\else $L_{\odot}$\fi}
\def\msolyr{\ifmmode {\,M_{\odot}\,{\rm yr}^{-1}}\else $M_{\odot}\,{\rm yr}^{-
1}$\fi}
\def\mdot{\ifmmode {\,\dot{M}}\else $\dot{M}$\fi}
\def\mdotyr{\ifmmode {\,\dot{M}\,yr^{-1}}\else $\dot{M}\,yr^{-1}$\fi}

\begin{document}

\title{ 
Long-Slit Observations of Extended C II $\lambda$1335 Emission Around 
V854~Centauri and  
RY~Sagittarii$^1$}

\author{Geoffrey C. Clayton$^2$ and T. R. Ayres$^3$}

\altaffiltext{1}{Based on observations obtained with the 
NASA/ESA Hubble
Space Telescope which
is operated by STScI for the Association of Universities for 
Research in 
Astronomy Inc. under
NASA contract NAS5-26555}

\altaffiltext{2}{Department of Physics and Astronomy, Louisiana State 
University, Baton Rouge, LA 70803; gclayton@fenway.phys.lsu.edu}
\altaffiltext{3}{Center for Astrophysics and Space Astronomy, 389-UCB,
University of Colorado,
Boulder, CO 80309-0389; ayres@casa.colorado.edu}

\begin{abstract} 
We have obtained long-slit far-ultraviolet (1150--1730~\AA)
spectra of the 
R~Coronae Borealis (RCB)
stars V854~Cen and RY~Sgr, near maximum light and pulsational 
phase zero,
with the Space Telescope Imaging Spectrograph (STIS) on {\it Hubble Space 
Telescope}\/ ({\it HST}). 
The far-UV spectrum of each star shows 
a photospheric continuum
rising steeply toward longer wavelengths, and a prominent emission
feature at \ion{C}{2} $\lambda$1335.  RY~Sgr displays a second, but fainter,
emission attributed to \ion{Cl}{1} $\lambda$1351 (which is radiatively
fluoresced by \ion{C}{2} $\lambda$1335), but \ion{Cl}{1}  is
weak or absent in V854~Cen.
Most surprisingly, the \ion{C}{2} emission of V854~Cen is significantly 
extended along the
slit by $\pm$2.5\arcsec, about 
$6\times 10^3$~AU at the
distance of the star.  The \ion{C}{2} feature
of RY~Sgr exhibits no such gross extension.  Nevertheless,
subtle broadenings of the \ion{C}{2} emissions beyond the
point response profile suggests inner clouds of radius $\sim$0{\farcs}1 (250~AU) 
around
both stars.  
V854 Cen is only the third RCB star after R CrB and UW Cen known to have a resolved shell.
\end{abstract}


\section{Introduction}
The R Coronae Borealis (RCB) stars are a small group of hydrogen-deficient carbon-rich 
supergiants which undergo spectacular declines in brightness
at apparently irregular intervals tied to their pulsation cycles (Clayton 1996).  
A cloud of 
carbon-rich dust forms along the line of sight, eclipsing the photosphere.
As the photospheric light is extinguished, a rich 
narrow-line ($\sim50$~km~s$^{-1}$) emission spectrum appears. 

In the visible, the narrow-line spectrum---referred to as E1---consists of 
many lines of 
neutral and singly 
ionized metals (e.g., Alexander et al.\ 1972).  
Most of the lines in the E1 spectrum
are short-lived, fading within two or three weeks. What remains is primarily a 
broad-line (BL: 100--200 km~s$^{-1}$) spectrum consisting of 
fewer lines.   
Some of the early-decline emissions---primarily 
multiplets of \ion{Sc}{2} and \ion{Ti}{2}, also narrow and referred to 
as E2---remain strong for an 
extended period of time; especially \ion{Sc}{2} (7) $\lambda$4246.
The E2 lines primarily are low excitation. 
The Balmer lines typically are very weak owing
to the hydrogen deficiency of 
the RCB stars, and do not go into emission except in the case of 
the relatively hydrogen-rich V854~Cen.
The late-decline visible BL spectrum is dominated by 
\ion{Ca}{2} H and K, 
the \ion{Na}{1} D lines, and \ion{He}{1} $\lambda$3888.  
\ion{He}{1} $\lambda$10830 also has been seen in R~CrB (Querci \& Querci 1978;
Rao et al. 1999).
The BL spectrum remains visible until the star  
returns to maximum light (when the BL emissions are
overwhelmed by the rejuvenated photospheric continuum).
A few optical nebular lines, such as [\ion{O}{2}], also
are seen in emission (Herbig 1949; Rao \& Lambert 1993).

The UV spectrum undergoes a similar evolution  
(Clayton et al.\ 1992a; Lawson et al.\ 1999).
The very early-decline spectrum (E1) is dominated by
blends of many emission lines 
which form a pseudo-continuum. 
The \ion{Mg}{2} doublet is present, but not particularly
strong during the E1 phase.    
The E2 spectra still show substantial blended emission, but \ion{Mg}{2}, 
\ion{Mg}{1} $\lambda$2852, and 
some of 
the \ion{Fe}{2} resonance lines have strengthened.   
The late-decline BL spectrum is thought to be characterized by blended 
emission from 
multiplets of \ion{Fe}{2} (2) $\lambda$2400, 
\ion{Fe}{2} (1) $\lambda$2600, \ion{Fe}{2} (62, 63) $\lambda$2750, 
as well as \ion{Mg}{2} and \ion{Mg}{1}.
\ion{C}{2} $\lambda$1335 and \ion{C}{3}] $\lambda$1909 also are 
seen, although their behavior is not as well documented; they
are thought to be in the BL class as well (Lawson et al.\ 1999). 
Unlike other emission lines, \ion{C}{2} $\lambda$1335 can be seen at all times,
perhaps thanks to its location deep in the UV where
the stellar continuum is faint.  

Here, we describe and compare high sensitivity far-UV observations of 
two key RCB stars---V854~Cen and RY~Sgr---obtained with the {\it HST}\/ 
Space Telescope Imaging Spectrograph
(Woodgate et al.\ 1998).  Our objective was
to extend the exploration of the dusty ejections
of these enigmatic objects to a spectral region where the photospheric
continuum is weak, and thus emissions from
the nebular environment of the stars could be
probed more directly.

\section{Target Stars}

RY~Sgr and V854~Cen are among the brightest of the RCB stars at maximum light.
On the one hand,
RY~Sgr is a relatively normal RCB star, with a helium/carbon atmosphere
(He/C\,$\sim$100) showing only trace amounts of hydrogen.  On the
other hand, V854~Cen is the least hydrogen-deficient RCB star, with
H/C$\sim$2, while the He/C ratio is 10--60, with the lower value the 
most likely (Asplund et al.\ 1998).
Perhaps related to the high hydrogen abundance,
the UV decline spectra of V854~Cen differ from those 
of  RY~Sgr (and R~CrB) in an important respect:
the presence of strong BL
emission from \ion{C}{2}] $\lambda$2325 and \ion{C}{1} $\lambda$2965
(Clayton et al.\ 1992a,b).

Previously, Holm \& Wu (1982) reported \ion{C}{2} $\lambda$1335 emission in 
an {\it IUE}\/ low-dispersion spectrum of RY~Sgr
at a level of $2\times$10$^{-14}$ 
ergs~cm$^{-2}$~s$^{-1}$.  
\ion{C}{2} 
has been detected 
in {\it IUE}\/ spectra of R~CrB and V854~Cen as well (Holm \& Wu 
1982; Holm et al.\ 1987; Clayton 1996).  Both of these stars 
have measured integrated fluxes
in C II
of $\sim8\times$10$^{-14}$ ergs~cm$^{-2}$~s$^{-1}$ (Brunner, Clayton, \& Ayres 
1998).
Lawson et al. (1999) reported integrated $\lambda$1335
fluxes of $\sim$1$\times$10$^{-13}$ ergs~cm$^{-2}$~s$^{-1}$ for
V854~Cen from 1991--1993.  At these flux levels, both RCB's would be
easy targets for the low-resolution G140L mode of STIS.

The distance to V854~Cen, assuming M$_V \sim -5$ mag 
and $E(B-V)\sim 0.\!^{\rm m}$1,
is $\sim$2.5~kpc
(Lawson et al.\ 1990; 
Alcock et al.\ 2001).  With the same assumptions, RY~Sgr is at a distance of
$\sim$1.5~kpc. 

\section{Observations}

Both RCB stars were observed near maximum light.
The RY~Sgr pointing was on 1998~April~25 (JD~2450929),
at $\phi=0.1$ of its 38-day pulsational cycle (ephemeris of 
Lawson \& Cottrell 1990).
We utilized
the G140L mode with the
52$\arcsec\times 0{\farcs}5$ slit, covering the spectral range 
1150--1730~\AA\ at $\sim$2~\AA\ resolution, with 
0{\farcs}1 resolution along the slit.  
Following a standard CCD acquisition, we 
obtained exposures of 1.97~ks and 2.65~ks
in consecutive orbits.
V854~Cen was observed on 1999~April~17 (JD~2451286),
at $\phi=0.9$ of its 43.2-day pulsational cycle (ephemeris of 
Lawson et al.\ 1992, 1999). 
We obtained exposures of 2.01~ks and 2.94~ks
in consecutive orbits.
The STIS spectrum of RY~Sgr was discussed previously
by Clayton et al.\ (1999a).

We reprocessed both sets of observations according to the prevailing STIS 
calibrations in late~2000, using
the on-the-fly facility of the {\it HST}\/ archive.
We co-registered each pair of 2-D long-slit spectral frames by measuring the 
cross-dispersion
(``$s$'') centroid of the spatial profile extracted from a continuum
region 1520$\pm$30~\AA; and the in-dispersion (``$\lambda$'') position by 
fitting the isolated \ion{C}{2}
$\lambda$1335 feature.  After co-adding the registered frames,
we determined the image background in two bands flanking the spectral trace, 
beginning
125~pixels (1~pixel = 0{\farcs}0244) on either side of the center, and extending 
for 75~pixels.  The
background levels appeared to be slightly asymmetric along the $s$ axis of the 
image, so we allowed for
a linearly sloping correction.  We extracted a 1-D spectrum 
for each $\lambda$ bin by accumulating the 
intensities in a $\pm$20~pixel ($\pm$0{\farcs}5)
band\footnote{This spatial interval should 
contain essentially all of the flux of a
normal point source.} centered on the $s$ centroid of the spectral stripe.

Figure~1 illustrates the 2-D and extracted 1-D spectra of V854~Cen and RY~Sgr.
In the spatially resolved images (lefthand panels), the central $\pm$0{\farcs}5 
is displayed on a linear scale, while the 
outer region has a logarithmic stretch to emphasize low-intensity 
spatial/spectral structure.  The faint extended
``peppering'' in both images near 1300~\AA\ represents 
enhanced residual photometric noise from the subtraction of the diffuse atomic
oxygen airglow feature (which fills the 0{\farcs}5 wide slit).  The extended 
\ion{C}{2} emission near
1335~\AA\ in the V854~Cen image is conspicuous.  In the 1-D spectral traces, the 
solid curves represent the extracted
fluxes, while the dashed lines are the 1\,$\sigma$ photometric noise levels.  
The jagged intervals near 1215~\AA\ are
residuals (mostly noise) from the subtraction of the bright atomic 
hydrogen airglow feature.  Note the logarithmic flux scale.

Both stars display prominent
\ion{C}{2} $\lambda$1335 emission, which dominates the spectrum below 
1400~\AA.  
RY~Sgr also shows, weakly, the
\ion{C}{3} $\lambda$1175 multiplet, and \ion{Cl}{1} $\lambda$1351 
(which likely 
is radiatively pumped by
\ion{C}{2} $\lambda$1335: see Clayton et al.\ 1999a).  However, V854~Cen lacks 
both features, at least at
the sensitivity levels of our STIS G140L spectroscopy.  
V854~Cen has a step-like energy
distribution longward of 1400~\AA\ (probably due to
the photoionization edges of \ion{C}{1} at 1349~\AA\ and 1444~\AA), 
punctuated by strong absorptions due to the 
\ion{C}{1} resonance multiplets
at 1560~\AA\ and 1657~\AA.  RY~Sgr has a more smoothly rising continuum 
distribution, with only weak
1657~\AA\ absorption apparent, flanked by \ion{Al}{1} $\lambda$1670.  
Additional 
absorption structure common to both stars is thought to
be due to carbon monoxide A--X
4th-positive system bands (Clayton et al.\ 1999a).
Integrated fluxes (or upper limits) 
of key spectral features are listed in 
Table~1.  In the case of \ion{C}{2} $\lambda$1335, the line fluxes were
integrated over the interval 1326--1344~\AA, after a small background
continuum level was removed (see Fig.~1). 
Because of the low resolution ($\sim$2 \AA), we cannot confirm from these 
data that the \ion{C}{2} $\lambda$1335 line is a broad line.

The STIS \ion{C}{2} flux for RY~Sgr is in good agreement with
the {\it IUE}\/ value reported by Holm \& Wu (1982).
However, the STIS measurement of the
point-like \ion{C}{2} emission in  V854~Cen is much smaller
than the {\it IUE}\/ fluxes published by Brunner, Clayton, \& Ayres (1998) and
Lawson et al.\ (1999).  The origin of the discrepancy can be traced to the
importance of the extended diffuse emission component, as will be
described shortly. 

Figure~2 compares the cross dispersion profiles of \ion{C}{2} $\lambda$1335 in 
the two
RCB stars.  The upper panels illustrate the behavior well beyond the stellar 
core.  The lower panels 
depict the near-star region.  Points and error bars are the observed spatial 
profiles
at 1335~\AA, while
the thin solid curves refer to the continuum band 1520$\pm$30~\AA\ 
(we would have used the continuum
in the immediate vicinity of the \ion{C}{2} feature, but it 
is too weak to yield a
suitable ``core'' profile).  V854~Cen exhibits smoothly-extended \ion{C}{2}
emission, limb-darkened rather than brightened, reaching out to about 
$\pm$2{\farcs}5.  No comparable diffuse extended 
feature is apparent around RY~Sgr.

However, as shown in the lower panels, both stars display broader \ion{C}{2} 
spatial profiles in the
near-star region than the continuum trace; and indeed the V854~Cen inner profile 
is offset from the stellar
continuum ``core'' by about 10~milliarcseconds (25~AU at the distance of the 
star).  
The continuum spatial profiles
of each star are the thicker solid curves in the lower panels.  The shaded 
profile represents a
$\lambda$1520 continuum trace
$\pm$1\,$\sigma$ based on an average of G140L 52\arcsec$\times$0{\farcs}5 
observations of several unresolved white dwarf (WD) stars (obtained from the {\it HST}\/ 
archive); 
the dot-dashed curve is the $\lambda$1335 trace from the same observations
(without errors, which are of similar magnitude to
the $\lambda$1520 example).  The $\lambda$1335 WD profile is slightly wider 
than the longer wavelength
trace at a level that is significant with respect to the standard error of the 
mean, but which hardly would be
noticeable in the comparisons, for example, in the upper panels.  (In the 
near-star examples, a background
was determined in the zone $\pm$(0{\farcs}5--2{\farcs}5), so that the influence 
of the extended emission
of V854~Cen would be suppressed, and all the stars---including the comparison 
WDs---would be treated 
on the same basis.)

\section{The Extended \ion{C}{2} Emission}

\subsection{The outer 2{\,\farcs}5}

The surface intensity of the extended \ion{C}{2} emission of V854~Cen has an 
average value of
$1.2\times 10^{-14}$~ergs cm$^{-2}$ s$^{-1}$ ($^{\prime\prime}$)$^{-2}$, having
compensated for $0.\!^{\rm m}$9
of absorption at 1335~\AA\ [assuming $E(B-V)\sim 0.\!^{\rm m}$1].  The 
appearance of the emission suggests that it comes
from a filled volume or thick shell; rather than a thin shell,
which would be strongly 
limb-brightened.  For the sake of argument, we assume that the extended emission
represents material ejected in a steady wind, with an $r^{-2}$ density
fall-off.  The relatively mild decline of the observed surface
intensity away from the star
suggests that the inner part of the volume--perhaps 50\% in radius--is 
evacuated, otherwise the \ion{C}{2} flux would fall off much more rapidly.

Although we have only a
0{\farcs}5 wide cut across the structure, we will assume that it
is a uniform disk on the sky of radius 2{\farcs}5.
Integrating the specific intensity over that area yields a total dereddened emission of 
$2.4\times 10^{-13}$~ergs cm$^{-2}$ s$^{-1}$, corresponding to $1.8\times 
10^{32}$~ergs s$^{-1}$ 
($L_{\mbox{\ion{C}{2}}}\sim 0.05\, L_{\odot}$) at the
$\sim$2.5~kpc distance of V854~Cen.  
Since the extended emission has a significantly larger 
total flux than the \ion{C}{2} point 
source, and would fall entirely within the {\it IUE}\/ 
10{\arcsec}$\times$20{\arcsec} large 
aperture, the spatially integrated
emission ($1.0\times 10^{-13}$~ergs cm$^{-2}$ s$^{-1}$ before the
reddening correction) is what should be
compared to the {\it IUE}\/ measures given previously, thereby accounting for 
the apparent discrepancy mentioned above. 

There are two general possibilities to explain the extended \ion{C}{2}
emission: (1) collisional excitation in a warm ($T\sim 3\times 10^4$~K), 
perhaps shock-heated,
gas; or (2) resonance scattering of the intrinsic stellar \ion{C}{2}
emission feature in
an optically thick evelope.  In the latter case, the gas could be
collisionally ionized (and therefore warm), or photoionized (and thus
potentially much cooler).  Since atomic carbon can be photoionized out
of the low-lying metastable 2p$^2$\,$^1$D state (1~eV above
ground; edge at 1239~\AA) by \ion{H}{1} Ly$\alpha$, and since
V854~Cen has considerably more hydrogen than typical RCB's, the
latter case is a very real possibility.

\subsubsection{Collisional excitation}

In the collisionally excited warm plasma case, we can infer the
emission measure 
($EM\equiv \int{n_{\rm e}^{2}\,dV}$) from the apparent flux of 
\ion{C}{2} radiation.
The emissivity of \ion{C}{2} at $T_{\rm m}= 10^{4.5}$~K is
$1.6\times 10^{-19}$~ergs cm$^{3}$ s$^{-1}$ (cf.,
Kaastra, Mewe, \& Nieuwenhuijzen [1996]; for a solar carbon to hydrogen ratio
of $\log{\epsilon_{\rm C}}= -3.48$).  We infer 
$EM\sim 1.1\times 10^{51}$~cm$^{-3}$ (assuming that all of the 
electrons come from carbon, and that
all of the carbon is singly ionized, i.e., $n_{\rm C}/n_{\rm e}\sim 1$).  The 
radius of the region is
2\farcs5, or $6\times10^{3}$~AU (0.03~pc) at the distance of V854~Cen. 

Given the thick-shell geometry described previously, and the assumption of
an $r^{-2}$ outward decline in density, one can relate
the average carbon density $<n_{\rm C}>$ (which equals the electron
density), and the total carbon number $N_{\rm C}$, to the
emission measure.  We find $<n_{\rm C}>\sim$1.2~cm$^{-3}$ (which corresponds
to an average particle density of $\sim$17~cm$^{-3}$ for
H/C= 2 and He/C= 10), and $N_{\rm C}\sim 1.6\times10^{51}$.  The mean
mass per carbon atom, with the abundances given above, is $9\times10^{-23}$~g;
so the total mass of material in the shell is about $1.4\times10^{29}$~g,
or $7\times 10^{-5}$~$M_{\odot}$.  To fill such a shell would require a
wind of 200~km~s$^{-1}$ (such as is seen in the gas
associated with dust formation in RCB stars) acting for a period of about
75~yr (to achieve the proposed $3\times10^3$~AU thickness of the structure);
the steady mass loss rate would have to be of order 
$\dot{M} \sim 1\times 10^{-6}\, M_{\odot}$~yr$^{-1}$.
The lifetime of the RCB evolutionary phase is uncertain,
but R~CrB, itself, has been an RCB star for at least 200~yr (Clayton 1996).  
If the
gas is cooler than the temperature of maximum \ion{C}{2} emissivity, the
inferred densities could increase substantially.  

A difficulty with the collisional excitation
model---which leads directly into consideration of the scattering case---is 
that the predicted column density of \ion{C}{2} is substantial, and
resonance absorption of the intrinsic stellar \ion{C}{2} emission feature
should be severe (line center optical depths of $10^2$, or so); unless
the majority of the envelope material is strongly blueshifted (by
more than the 200 km~s$^{-1}$ line width, if \ion{C}{2} is a member of
the broad-line class) and the internal velocity dispersion is low
($<50$~km s$^{-1}$).  That would require the envelope material to be
suddenly accelerated to its terminal velocity, which seems unlikely.  

\subsubsection{Photoexcitation}

In the scattering case, we imagine that atomic carbon gas streaming
away from the star, perhaps dragged along with radiatively accelerated
dust grains, becomes photoionized by a flux of \ion{H}{1} Ly$\alpha$
radiation from the star (produced by the same ``chromospheric'' processes
that account for the intrinsic \ion{C}{2} emission), creating a
C$^+$ Stromgren sphere.  The C$^+$ ions, in turn, could resonantly absorb
the \ion{C}{2} $\lambda$1335 multiplet emission from the star; strongly
attenuating the stellar emission and redistributing it throughout the
Stromgren sphere by multiple scatterings.  The process would be nearly
conservative because the envelope density is so low that collisional
deactivations would be negligible.  Since the extended $\lambda$1335
emission of V854~Cen 
appears to be about 7$\times$ the flux of the stellar point source, we
infer that the average optical depth across the stellar \ion{C}{2} features
must be $\sim$2.  If we imagine that the flow smoothly accelerates away
from the star up to a terminal velocity of --200~km s$^{-1}$, such that each
velocity bin has equal \ion{C}{2} column density, the optical depth at
each point across the stellar emission feature will be 
$\tau_{\rm lc}\sim 10^{-13}\, \frac{\pi\,e^2}{m\,c^2}\, f\, \lambda_{0}\,
v_{\rm max}^{-1}\, N_{\rm C\,\footnotesize II}$, where the 
line oscillator strength $f$ is 0.1 for the two strong transitions of
the multiplet, the line center
wavelength $\lambda_{0}$ is in \AA, the flow terminal velocity $v_{\rm max}$
is in km~s$^{-1}$, and  $N_{\rm C\,\footnotesize II}$ is the column density
of C$^+$ ions in the ground state (cm$^{-2}$).  For the indicated
$\tau_{\rm lc}\sim 2$, one infers a column density of about $1\times 10^{15}$
cm$^{-2}$.
If the shell thickness is $3\times 10^3$ AU, then the average C$^+$ density
would be about 0.02~cm$^{-3}$, considerably smaller than we found for
the collisionally excited case ($\sim 1$~cm$^{-3}$).  The mass-loss
requirements for
filling the volume would be reduced dramatically as well.  

The one caveat
is that since the Ly$\alpha$ photoionization of carbon  occurs from an
excited state, the local population of susceptible atoms depends very
sensitively on the temperature (unlike a normal \ion{H}{2} region where
the photoionization is from the ground state, and locally virtually every
H atom can be ionized, if sufficient hard photons are present).  In a
partially ionized carbon envelope, the local density could be significantly
higher than indicated by the \ion{C}{2} resonance absorption.  We do see
strong \ion{C}{1} $\lambda$1560 and $\lambda$1657  resonance absorptions in
the point source spectrum of V854~Cen, but the extent to which these are
intrinsic to the stellar photosphere or contributed by the shell cannot be
established with these low-resolution traces.  There do not appear to be
any sensible extended \ion{C}{1} emission zones at those wavelengths in the
spectral image,
however, suggesting that the carbon probably is mostly ionized in the 
2\farcs5 shell; but the faintness of an extended \ion{C}{1} region might simply
be due to strong photospheric \ion{C}{1} absorption that suppresses the
photon flux available to excite the neutral gas.

Undoubtedly, one of the key differences between V854~Cen and RY~Sgr, as far
as revealing an extended carbon envelope, is
the large hydrogen abundance of the former: the associated strong
Ly$\alpha$ emission might be instrumental in ionizing the envelope so that
it would be susceptible to illumination by the stellar \ion{C}{2} 
emission.  On the other hand, the fact that both stars apparently possess inner  
\ion{C}{2} 
emission zones might indicate that the near-star regions are collisionally
ionized and excited, rather than photo-dominated.

Many of these issues could be addressed more definitively with higher spectral
resolution measurements, which, for example, could disentangle Doppler shifted 
shell
absorptions, if present, from the intrinsic stellar emission features.

\subsection{The inner 0{\,\farcs}1}

We can obtain estimates for the inner emission zone of V854~Cen.  We
fitted a simple model to the apparent cross-dispersion profile of the \ion{C}{2}
feature, using the $\lambda$1520 continuum profile as an estimate of the
line spread function at 1335~\AA.  The model consisted of a point source, 
and two uniform surface brightness disks,
one 0{\farcs}1 in radius, the other 0{\farcs}3.  The best fit was achieved by
shifting the point source by 0{\farcs}01 relative to the
continuum ``core,'' the smaller disk by 0{\farcs}02 in
the same direction, and the larger disk by 0{\farcs}02 in the opposite
direction.  The relative fluxes were in the ratio 1.0\,:\,1.0\,:\,0.75.  The
innermost disk has a surface brightness about 40~times that of the 2{\farcs}5
extended diffuse component.

At the distance of V854~Cen, the inner 0{\farcs}1 disk has a radius of about
250~AU, and the average excess \ion{C}{2} flux, corrected for reddening, 
is $1.3\times 10^{-14}$~ergs cm$^{-2}$ s$^{-1}$.
The corresponding \ion{C}{2} luminosity is
$1\times 10^{31}$~ergs s$^{-1}$, and the carbon emission measure is 
$EM_{\rm C}\sim 6\times 10^{49}$~cm$^{-3}$ 
(again assuming the collisional excitation model in
which carbon is the major electron donor, and is mostly singly 
ionized).  Unfortunately, converting the emission measure into an 
average carbon
density is more difficult in this situation, because the inner radius of
the zone could be quite small (i.e., $\sim R_{\star}$):  For a 
mass-conservative $r^{-2}$ density law (for a constant velocity outflow), the 
average density scales as 
$(r_{\rm min}/r_{\rm max})^{1/2}$, if the inner radius $r_{\rm min}$ is
much smaller than the maximum radial extent of the zone.  

We can, however, 
set an upper limit to the average density by assuming that the region is
spherical with uniform density.  In that case, 
$<n_{\rm C}>\sim$17~cm$^{-3}$ (yielding
an average particle density of $\sim$240~cm$^{-3}$ for
H/C= 2 and He/C= 10).  The total mass of material in the volume
would be about $3\times10^{26}$~g,
or $2\times 10^{-7}$~$M_{\odot}$, where again this would be an upper
limit.  The crossing time for a 200~km~s$^{-1}$
wind would be about 7 yr, so the upper limit to
the steady mass loss rate would 
be $\dot{M} < 3\times 10^{-8}\, M_{\odot}$~y$^{-1}$, significantly
smaller than the mass flux required to fill the outer shell.

Similar intensity and emission measure
estimates would apply to the inner emission zone
of RY~Sgr (which could be fitted adequately using only the point source
plus a 0{\farcs}1 radius inner disk).  However,
the He/C abundance ratio could be as much as
$10\times$ that of V854~Cen, leading to a corresponding order of magnitude
increase in the total particle density and 
mass of the emitting material.

\section{Discussion}

The BL lines such as \ion{Ca}{2} and \ion{Na}{1} require relatively 
modest energies ($\lesssim6$ eV) for ionization and excitation.
But, other features such as \ion{C}{2} $\lambda 1335$, \ion{C}{3}] 
$\lambda 1909$ and \ion{He}{1} $\lambda$10830 require significantly higher 
energies.
The ionization potentials are 11 and 24~eV, respectively, to C$^+$ and C$^{++}$. 
The lower level of the 
\ion{He}{1} $\lambda$10830 transition is 20~eV above ground.  
It is metastable and could be populated from below by collisions
or from above by cascades following recombination 
from \ion{He}{2} (Geballe et al.\ 2001).
Since the RCB stars discussed here have only modest photospheric
temperatures, $T_{\rm eff} = 6000$--7000~K, they cannot produce 
\ion{He}{2} through photoionization.
At the same time, it is likely that there are enough UV photons to singly ionize 
carbon (Rao \& Lambert 1993), particularly from its low-lying
2p$^2$\,$^1$D and 2p$^2$\,$^1$S states (with
edges at 1239~\AA\ and 1444~\AA, respectively).

Rao et al.\ (1999) suggested that the high-excitation lines arise in an 
accretion disk around a white dwarf secondary.  
In that model, the formation zone for the hot-broad lines would
be relatively small ($\lesssim 3$~AU) assuming $n_{\rm e}\sim 10^7$~cm$^{-3}$,
and offset a few AU from the primary.  
The apparent displacement
of the \ion{C}{2} emitting region by 0{\farcs}01 or 25 AU, 
seen in V854 Cen,  fits that model in a qualitative way.
However, the 
fact that the majority of the \ion{C}{2} $\lambda$1335 emission of
V854~Cen arises in an outer diffuse zone thousands of AU from the star 
renders the WD model less appealing.
Collisional excitation in the extended thick shell perhaps is more plausible.  
High velocities (up to $\sim$400~km~s$^{-1}$) are seen in 
absorption and emission lines during declines, probably originating 
from gas dragged 
along with the dust blown 
away from the star by radiation pressure 
(Rao \& Lambert 1993; Feast 1996, 2001; Clayton 1996).  
Rao \& Lambert (1993) suggest that V854 Cen may have a high velocity
bipolar outflow.
In these fast-moving clouds, 
excitation might take place through atomic collisions or shocks (Feast 2001). 

An additional curiosity of the BL region are the so-called
``decline eclipses.'' 
The BL lines often are constant throughout declines, 
but sometimes are seen to fade significantly over
$\sim$100 days indicating that
a portion of the BL emission is coming from
within a few AU of the star (Herbig 1949; Clayton et al.\ 1992a; 
Lawson et al. 1999).
The extended \ion{C}{2} emission regions reported here bear some resemblance
to the BL region deduced for V854~Cen by Rao \& Lambert (1993) using
forbidden lines.  
Assuming a homogeneous spherical nebula with $n_{\rm e}$ = $n_{\rm 
p}$ = 50~cm$^{-3}$ 
they infer r$_{\rm neb}\sim$ 
930~AU for V854~Cen at a distance of 2.5~kpc. If the filling factor is $f=0.1$, 
then  r$_{\rm neb}\sim$ 
$2.1\times 10^3$~AU.  Our results are complementary in the sense that
while we have the advantage of directly measuring the apparent size
of the emission region(s) on the sky, we lack a diagnostic to 
establish the mean density, and thus the filling factor if the material
is highly clumped.  Unfortunately,
a similar comparison cannot be applied to RY~Sgr,
because the same range of decline spectra do not exist.

Although, it is difficult to make definitive statements regarding the 
absence of an extended emission region around RY~Sgr, we point out the 
following:  On the one hand,
RY~Sgr is believed to be less distant than V854~Cen so a similar emission region 
would appear to be even more extended. Such an envelope around RY~Sgr,
with the same mass content
as that of V854~Cen, would have as little as one tenth the
carbon (owing to the larger He/C ratio),
and thus only {\it one hundredth}\/ the potential \ion{C}{2} emission measure
(if carbon is the major electron donor).  
Alternatively, the \ion{C}{2} envelope
of V854~Cen might be ionized into visibility by a significant \ion{H}{1} 
Ly$\alpha$ emission flux,
whereas the \ion{H}{1} resonance emission of RY~Sgr would be much less 
prominent, owing to the severely
reduced hydrogen abundance.

On the other hand, perhaps an extended low density BL region simply is not 
present around RY~Sgr. 
V854~Cen arguably is significantly
different from RY~Sgr---and the majority of RCB 
stars---in its hydrogen and other element abundances,
its high level of decline activity, and its emission line spectrum 
(Clayton 1996, Asplund et al.\ 2000).  So, perhaps it would not be surprising 
if it differs in this manner as well.

The behavior of the V854~Cen lightcurve over the last century provides some support 
for the finding in
{\S}4.1 that the 2{\farcs}5 shell likely is filled in the outer 50\% and 
of lower density
in the inner part.  Although V854~Cen is one of the most active RCB
stars today, it had an even higher mass loss rate 
in the period of 1913--1952 when it was 
generally fainter than 13th magnitude (Clayton 1996, and references therein).  
The shell could be filled by gas traveling at 200 km~s$^{-1}$ in $\sim$75 yr,
which 
matches well with the observed timescales of the decline activity in V854~Cen. 
Approximately 50~yr of high activity followed by 50~yr of lesser activity 
could result in a relatively greater density in the outer part of the shell.

Other than V854~Cen, the only resolved RCB spatial structures are the large
IRAS shell of R~CrB (Gillett et al.\ 1986), and a tenuous $\sim$7\arcsec~radius 
envelope around
UW~Cen seen in dust-reflected visible light (Pollacco et al.\ 1991; 
Clayton et al.\ 1999b).  
At the assumed---but very uncertain---distance of UW~Cen, 
the radius would be $\sim$0.2~pc (4$\times 10^{4}$ AU).  
It has been suggested that the UW~Cen nebula is a fossil planetary
nebula shell created by a slow-moving
outflow from the star over thousands of years (Clayton et al.\ 1999b).
But, in analogy with V854~Cen, such a shell also might be produced 
by high velocity ejections during the present RCB-star epoch.

\acknowledgments
This work was supported by STScI grant GO-07477.01-96A and NASA grant
NAG5-3226 (TRA).
We thank Denise Taylor, Steve Hulbert 
and Brian Espey for their help in obtaining and reducing the data 
reported in this paper.

\clearpage
\begin{table*}
\normalsize
\caption{Observed Fluxes at Earth from STIS Spectra}
\begin{tabular}{cccc}
\tableline
\tableline
Star & \ion{C}{3} 1175	& \ion{C}{2} 1335 & \ion{Cl}{1} 1351\\
  & \multicolumn{3}{c}{($10^{-15}$~ergs cm$^{-2}$ s$^{-1}$)}\\
\tableline
V854~Cen       & $<$0.5  & 14.9$\pm$0.3$^1$  & $<$0.2 \\
RY~Sgr         &  1.1$\pm$0.2  & 23.5$\pm$0.4   & 1.2$\pm$0.1 \\
\tableline
\end{tabular}
\tablenotetext {1}{This value includes flux from $\pm$0{\farcs}5 from the star. The integrated flux 
including the extended emission is significantly higher. See text.}
\end{table*}

\begin{figure*}
\caption{2-D (lefthand panels) and 1-D (righthand panels) 
spectra of V854~Cen (upper) and RY~Sgr (lower).}
\end{figure*}

\begin{figure*}
\plotone{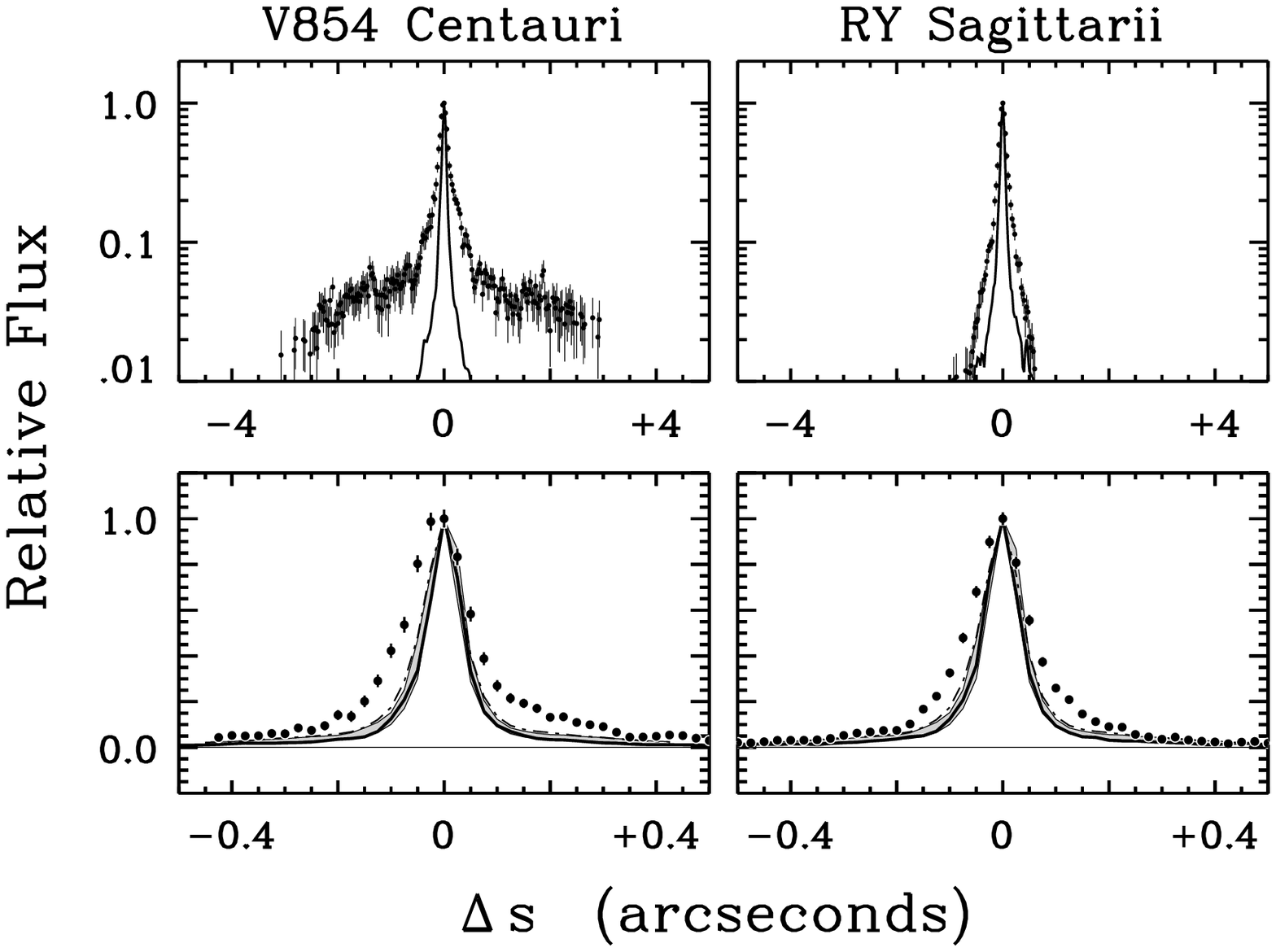}
\caption{Comparison of spatially extended cross dispersion profiles of the 
\ion{C}{2} 
$\lambda$1335 feature in the two
RCB stars.}
\end{figure*}


\begin{references}

\def\same{ \rule[.5ex]{2em}{.4pt} }

\reference{} Alcock, C. et al. 2001, ApJ, 554, 298

\reference{} Alexander, J. B., Andrews, P. J., Catchpole, R. M., Feast, M. W., 
Lloyd 
Evans, T., Menzies, J. W., Wisse, P. N. J., and Wisse, M. 1972, MNRAS 158, 305

\reference{} Asplund, M., Gustafsson, B.,  Rao, N.K., \& Lambert, D.
L. 1998, A\&A, 332, 651

\reference{} Asplund, M., Gustafsson, B., Lambert, D.
L., \& Rao, N.K. 2000, A\&A, 353, 287

\reference{} Brunner, A., Clayton, G.C, \& Ayres, T.R. 1998, PASP, 110, 1412 

\reference{} Clayton, G.C., Whitney, B.A., Stanford, S.A., \& 
Drilling, J. S. 1992a, ApJ, 397, 652

\reference{} Clayton, G.C., Whitney, B.A., Stanford, S.A.,
Drilling, J. S., \& Judge, P.G. 1992b, ApJ, 384, L19

\reference{} Clayton, G.C. 1996, PASP, 108, 225

\reference{} Clayton, G.C., Ayres, T.R., Lawson, W.A., Drilling, J.S., Woitke, 
P., \& Asplund, M. 1999a, 
ApJ, 515, 351 

\reference{} Clayton, G.C., Kerber, F., Gordon, K.D., Lawson, W.A, Wolff, M.J., 
Pollacco, D.L., and Furlan, E. 
1999b, ApJ Letters, 517, L143 

\reference{} Feast, M. 1996,  {\rm A.S.P. Conf. Ser.}, 96, 3

\reference{} Feast, M. 2001, in Eta Carina and Other Mysterious Stars, ASP Conf. 
Ser., in press

\reference{} Geballe, T.R., Evans, A.E., Smalley, B., Tyne, V.H., \& Eyres, 
S.P.S. 2001, astro-ph/0102043

\reference{} Gillett, F. C., Backman, D. E., Beichman, C., and Neugebauer, G. 
1986, ApJ, 310, 842

\reference{} Herbig, G.H. 1949, ApJ, 110, 143

\reference{} Holm, A. V., \& Wu, C. C. 1982, in Advances in 
Ultraviolet Astronomy: Four Years of IUE Research, NASA CP-2238, p. 429

\reference{} Holm, A. V., Hecht, J., Wu, C. C., \& Donn, B. 1987, 
PASP, 99, 497

\reference{} Kaastra, J.\ S., Mewe, R., \& Nieuwenhuijzen, H.\ 1996, in 
UV and X-ray Spectroscopy of Astrophysical and Laboratory Plasmas,
Eds.\ K.\ Yamashita and T.\ Watanabe 
(Tokyo: Universal Academy Press), p.\ 411

\reference{} Lawson, W. A., \& Cottrell, P. L. 1990, MNRAS,  242, 259

\reference{} Lawson, W. A., Cottrell, P. L., Kilmartin, P. M., and Gilmore, A. 
C. 1990, 
MNRAS, 247, 91

\reference{} Lawson, W.A.,Cottrell, P.L., Gilmore, A.C., \& Kilmartin, 
P.M. 1992, MNRAS, 256, 339

\reference{} Lawson, W.A., Maldoni, M.M., Clayton, G.C., Valencic, L., Jones, 
A.F., Kilkenny, D., van Wyk, F., 
Roberts, G., \& Marang, F. 1999, AJ, 117, 3007 

\reference{} Pollacco, D. L., Hill, P. W., Houziaux, L., and Manfroid, J. 1991, 
MNRAS, 248, 572

\reference{} Querci, M., \& Querci, F. 1978, A\&A, 70, L45

\reference{} Rao, N. K., and Lambert, D. L. 1993, AJ, 105, 1915

\reference{} Rao, N. K., et al. 1999, MNRAS, 310, 717 


\reference{}  Woodgate, B.\ E., et al.\ 1998, PASP, 110, 1183

\end{references}
\end{document}